\documentstyle[12pt,titlepage,epsfig]{article}

\renewcommand{\theequation}{\thesection.\arabic{equation}}
\def\NPB{{\em Nucl. Phys.} B}
\def\PLB{{\em Phys. Lett.}  B}
\def\PRL{{\em Phys. Rev. Lett. }}
\def\PRD{{\em Phys. Rev.} D}
\def\MPL{{\em Mod. Phys. Lett.}  A}
\def\CQG{{\em Class. Quant. Grav. }}
\def\laq{\raise 0.4ex\hbox{$<$}\kern -0.8em\lower 0.62
ex\hbox{$\sim$}}
\def\gaq{\raise 0.4ex\hbox{$>$}\kern -0.7em\lower 0.62
ex\hbox{$\sim$}}

\def\L{{\cal L}}

\def\half{\hbox{\small $\frac{1}{2}$}}
\def\beq{\begin{eqnarray}}
\def\eeq{\end{eqnarray}}
\def\be{\begin{equation}}
\def\ee{\end{equation}}
\newcommand{\figref}[2]{Fig. \ref{#1}#2}
\def\pbd{{\dot {\bar \phi}}}
\def\pbdd{{\ddot {\bar \phi}}}
\newcommand{\varby}[2]{{{\delta {#1}} \over {\delta {#2}}}}
\newcommand{\partby}[2]{{{\partial {#1}} \over {\partial {#2}}}}
\newcommand{\parttbyo}[2]{{{\partial{^2} {#1}}
   \over {\partial {#2}{^2}}}}
\newcommand{\parttbyt}[3]{{{\partial{^2} {#1}} \over
  {\partial {#2} \partial {#3}}}}
\newcommand{\bfig}{\begin{figure}}
\newcommand{\efig}{\end{figure}}
\newcommand{\bigepsfour}[1]{\begin{center}\epsfig{file=#1,width=17cm,
                        height=10cm,bbllx=60,bblly=40,
                        bburx=560,bbury=470}\end{center}}
\newcommand{\bigepstwo}[1]{\begin{center}\epsfig{file=#1,width=17cm,
                        height=6cm,bbllx=60,bblly=220,
                        bburx=560,bbury=470}\end{center}}

\begin{document}
\titlepage
\begin{flushright}
BGU-PH-98/12 \\
CERN-TH/98-411 \\
\end{flushright}
\vspace{22mm}
\begin{center}
{\bf {\Large Classical Corrections in String Cosmology}}\\

\vspace{10mm}
\centerline{Ram Brustein}
\medskip
\centerline{\it {Department of Physics, Ben-Gurion
University, Beer-Sheva 84105, Israel}}
\bigskip
\centerline{Richard Madden}
\medskip
\centerline{\it {Theory Division, CERN, 1211 Geneva 23, Switzerland}}
\vskip 2 cm

{\large  Abstract} 

\end{center} 

\noindent
An important element in a model of non-singular string cosmology 
is a phase in which classical corrections saturate the growth of
curvature in a deSitter-like phase with a linearly growing dilaton 
(an `algebraic fixed point'). As the form of the classical corrections 
is not well known, here we look for evidence, based on a suggested 
symmetry of the action, scale factor duality and on conformal 
field theory considerations, that they can produce this saturation.
It has previously been observed that imposing scale factor duality on the
$O(\alpha')$ corrections is not compatible with fixed point behavior.
Here we present arguments that these problems persist to all orders
in $\alpha'$. We also present evidence for the form of a solution to 
the equations of motion using conformal perturbation theory, examine
its implications for the form of the effective action and find novel
fixed point structure. 

\vspace{5mm}

\vfill
\newpage

\renewcommand{\theequation}{1.\arabic{equation}}
\setcounter{equation}{0}
\section {Introduction}

Einstein's Theory of General Relativity precipitated a revolution
in cosmology, predicting in a quantifiable way a dynamical universe.
However, extrapolating this evolution back to very early times exposes
a number of puzzles. Firstly, to evolve to the universe 
we observe today the
early universe must be unnaturally smooth and flat. These problems
can be solved by introducing an early phase of accelerated evolution, 
generically known as inflation. But, secondly, beyond
these apparent fine tuning problems, the universe as we know it must
have a singularity of infinite density in the past \cite{hp}, indicating
that General Relativity itself is incomplete.

Our best current candidates for a unified theory of gravity and all 
other interactions are the various string theories. These theories generically 
predict that in addition to the
fields of the metric tensor, gravity contains a scalar component 
called the dilaton, whose vacuum expectation value
also controls masses of particles and 
the strength of the various gauge couplings. While the presence of a 
light dilaton was actually found to be damaging to standard forms of 
inflation based on the potential energy of a scalar field \cite{baddil},
a radically different scenario was proposed \cite{gv1,gv2} in which
the kinetic energy of the dilaton field drives inflation. Further, 
the natural origin for such a phase lies in 
the perturbative domain of string
theory, a weakly coupled, very flat universe. Objections have
been raised that this origin is itself a form of fine-tuning,
and while the situation is not yet completely clear, some aspects 
of these objections have been answered or spawned interesting
new speculations on the question of initial conditions. 
Nonetheless, this scenario,
dubbed the `pre-big-bang', utilizes the dilaton as 
a uniquely natural inflationary 
candidate, sidesteps the ambiguities inherent in placing the origin
of the universe in a near singular state and most interestingly leads to
possible observable signals \cite{signal}.

The scenario begins with the observation that solutions to the
lowest order equations of motion for the metric-dilaton system
come in duality pairs related by a  symmetry of 
string theory in cosmological backgrounds, scale factor duality (SFD). 
They consist of an inflationary branch in which
the Hubble parameter increases in time and which ends in a singularity
(the (+) branch) and a decelerated branch with decreasing Hubble
parameter which begins in a singularity (the ($-$) branch).  
The `pre-big-bang' scenario consists of a 
essentially empty universe beginning on a (+) branch
which evolves to high curvature and rapid expansion but 
instead of going all the way to a singularity, turns around and
decelerates into a ($-$) branch. This ($-$) branch
can then be smoothly joined with a standard radiation dominated 
Friedmann-Robertson-Walker (FRW) universe with constant dilaton, 
necessary to be compatible with observational limits on the variation
of gauge couplings and masses, at least from the time of nucleosynthesis
\cite{fixdil}. This combination provides a realistic cosmology
in which the high curvature phase joining the two branches, which
we expect to be accompanied by copious particle production, 
is identified with the `big bang' of the standard model.

To produce this joining of the two branches 
we will need to consider correction terms to the 
lowest order action which may allow exit from the (+) branch inflationary 
phase. This has proved a frustrating enterprise, leading to this
being called the `graceful exit problem'. In \cite{bm1} we generalized
earlier specific `no-go' theorems \cite{kmo} to show that the property
required of the additional sources is the ability to violate the 
Null Energy Condition (NEC) ($\rho+p \le 0$, where $\rho$ is the effective
energy density of the additional sources 
and $p$ is the pressure). On the negative side this rules out
standard sorts of sources such as potentials, other scalar fields,
perfect fluids with reasonable equations of state, etc. On the
affirmative side this points in the direction of quantum corrections 
which are known to be capable of violating such energy conditions.

Corrections to the lowest order action take the form of a dual series
in two expansion parameters. The first is the string length scale
$\lambda_s=\sqrt{\alpha'}$. Corrections in this parameter 
become important in the regime of 
large curvature. These are classical corrections related to the 
finite string size and are expected to play a role 
in regulating curvature growth.
The second is the dimensionless string coupling $e^{\phi}=g{^2}{_{string}}$, 
where $\phi$ is the dilaton expectation value. These are genuinely 
quantum corrections
since the power of $e^{\phi}$ counts the number of loops in the 
string worldsheet topology and they can, in principle, violate NEC.

We recently presented an explicit model of a graceful exit \cite{bm2},
following a suggestion \cite{gmv}, that $O(\alpha')$ classical correction
could limit curvature growth, leaving the universe in a de-Sitter like
phase (a `fixed point') with constant Hubble parameter
but a linearly growing dilaton.
Since the dilaton controls the strength of quantum loop corrections,
they will become stronger, eventually providing the source of 
NEC violation to complete the exit. 

While it is possible the loop corrections could accomplish the
graceful exit on their own, it appears to be difficult to tune
the theory to accomplish this. The `fixed point' behavior  seems to be necessary
to bridge naturally between the inflationary (+) branch and the
graceful exit. With a fixed point to rest in,  changes in the initial conditions
only  change the value of the dilaton at the time of entry to the fixed point,
but do not affect the behavior of the final exit phase.

Although the existence of such a fixed point is a question that
should be answerable from first principles, 
as we explain in detail in the next section, our knowledge of the form
of these corrections of higher order in $\alpha'$ is very limited.
Previous works simply selected corrections that exhibited fixed
point behavior from a family of corrections compatible with
the few known properties of these corrections. But not all members
of this family exhibit this behavior. In this work we discuss evidence
that classical corrections of higher order may or may not assist
the graceful exit by exhibiting attractive fixed points, and find evidence for
the position of the fixed point.

\renewcommand{\theequation}{2.\arabic{equation}}
\setcounter{equation}{0}
\section {Effective string cosmology}

\subsection {General Considerations}

String theory effective action takes the following form,
\beq
S_{eff}&=&
 \frac{1}{16 \pi \alpha'}\int d^4 x \sqrt{-g}\left[ e^{-\phi}
\left(\L_{0}+
\half \L_c(\phi,g_{\mu\nu} \ldots)\right)+\ldots \right]
\label{effacts} \\
\L_{0} &=& R+\partial_\mu\phi \partial^\mu\phi,
\eeq
where $g_{\mu\nu}$ is the 4-d metric and $\phi$ is the dilaton, 
the effective action is written here in the string frame.
$\L_c$ contains the corrections to the lowest order 4-d action
coming from a variety of sources, but here we restrict ourselves
corrections that are tree-level
in the string worldsheet, terms made up of
covariant combinations of the massless fields and their derivatives
(the graviton $g_{\mu \nu}$, the dilaton $\phi$ and the antisymmetric
tensor field strength $H_{\mu \nu \sigma}$, which we here set to
zero). As higher order corrections, they take the form of a series
expansion with expansion parameter $\alpha'=\lambda_s^2$, where
$\lambda_s$ is the string length scale.

We are interested in solutions to the equations of motion derived from
the action (\ref{effacts}) of the FRW type with vanishing spatial curvature
$ds^2= -n^2(t) dt^2+a^2(t) dx_i dx^i$ and $\phi=\phi(t)$. We include
the corrections in the form of their energy momentum
tensor $T_{\mu\nu}=\frac{1}{\sqrt{-g}}
\frac{\delta \sqrt{-g} e^{-\phi} \L_c}{\delta g^{\mu\nu}}$,
which will have the form $T^\mu_{\ \nu}=diag(\rho,-p,-p,-p)$. In addition 
we have another form of source term arising from the variation by $\phi$  
equation, $\Delta_\phi\L_c=\half \frac{1}{\sqrt{-g}}
\frac{\delta \sqrt{-g} e^{-\phi} \L_c}{\delta\phi}$.
 
In terms of these sources the equations of motion are 
\begin{eqnarray} 
3H^2+\half\dot\phi^2-3 H\dot\phi&=&
\half e^{\phi} \rho \label{n00eq} \\ 
-2 \dot H -3 H^2 + 2 H\dot\phi - \half \dot\phi^2 + \ddot\phi&=&
\half e^{\phi} p \label{n11eq} \\  
3 \dot H + 6 H^2 - 3 H\dot\phi + \half \dot\phi^2 - \ddot\phi&=&
\half e^{\phi} \Delta_\phi\L_c \label{phieq} \\  
\dot\rho+3 H(\rho+p)&=&
-\Delta_\phi\L_c \dot\phi, \label{nconseq} 
\end{eqnarray} 
$H=\dot a / a$.
The explicit $\phi$
dependence in these equations is an artifact of our attempt to
maintain consistency with earlier works. For the tree level
classical corrections $\rho$ itself will be
of the form $e^{-\phi} (\ldots)$. So the
corrections $\L_c$ will appear in the equations of
motion as polynomials
in $H$ and $\dot \phi$ and possibly higher derivatives.
 
Our knowledge of the form of these corrections is incomplete.
Efforts to fix them by requiring the action reproduce the
string theory S-matrix elements \cite{cor} can determine only some
of the coefficients of potential covariant terms in the action since
others do not contribute to the S-matrix or 
make contributions which overlap in form with those of other terms
\cite{cor2}. For example, in \cite{cor} they fix the contribution
\beq
\L_c={{k \alpha'}\over{2}} (\rho_0 R^{\mu \nu \lambda \sigma}
     R_{\mu \nu \lambda \sigma}+\rho_1 (\nabla \phi)^4), 
\label{tseytlin}
\eeq
and find $\rho_0=1$ and $\rho_1=0$, where $k=1,1/2$ for the bosonic and
heterotic string respectively (for the type II string $k=0$ and the
corrections start at higher order). We will thus find it convenient 
to fix our units such that $k \alpha'=1$. 
There are also determinations of
other contributions containing the antisymmetric tensor field strength.
Our knowledge of higher order corrections fades rapidly with 
increasing order.

But even (\ref{tseytlin}) is ambiguous, as we can make modifications to
this correction (`field redefinitions') of the form,
\be
S_{eff} \rightarrow S_{eff}+S_{mod}
\ee
\be
S_{mod}=
 \frac{1}{16 \pi \alpha'}\int d^4 x
\left( \varby{\sqrt{-g} e^{-\phi} \L_{0}}{{g_{\mu \nu}}} {\delta {g_{\mu \nu}}}
       + \varby{\sqrt{-g} e^{-\phi} \L_{0}}{\phi} {\delta \phi}
\right).
\label{redef}
\ee
We have added to the action factors consisting of the
lowest order equations of motion multiplied by
${\delta {g_{\mu \nu}}}$ and $\delta \phi$ which we will chose to
be explicitly proportional to $\alpha'$.
Since the corrections to
the lowest order equations of motion are also of order
$\alpha'$ for dimensional reasons, this is consistent
on the level of a truncated perturbation
expansion in powers of $\alpha'$. 
More importantly, these modifications are justified on the basis of the
fact they don't alter the on-shell scattering S-matrix and so have equal
standing with correction (\ref{tseytlin}) \cite{cor2}.
As we shall see, in spite of this equivalence they represent quite different 
effective actions for evolving cosmologies.

Explicitly (see \cite{mag,odd}) this becomes,
\beq
S_{mod}=
 \frac{-1}{16 \pi \alpha'}\int d^4 x \sqrt{-g} e^{-\phi}
\biggl\{ (R^{\mu \nu}+\nabla^{\mu} \nabla^{\nu} \phi
        -\frac{1}{2} g^{\mu \nu}(R+2 \nabla^2 \phi-(\nabla \phi)^2))
        {\delta {g_{\mu \nu}}} + \nonumber 
\eeq
\beq
        + 2 (R+2 \nabla^2 \phi-(\nabla \phi)^2) {\delta \phi} \biggr\},
\eeq
where we put
\beq
\delta g_{\mu \nu} &=& k \alpha' ( a_1 R_{\mu \nu}+
   a_2 \nabla_{\mu} \phi \nabla_{\nu} \phi+
   a_3 g_{\mu \nu} (\nabla \phi)^2+
   a_4 g_{\mu \nu} R+
   a_5 g_{\mu \nu} \nabla^2 \phi ) \\
\delta \phi &=& k \alpha' ( b_1 R +b_2 (\nabla \phi)^2+b_3 \nabla^2 \phi ).
\eeq
Explicitly evaluating the correction, we find it can be expressed
in terms of the following tensor structures (after some integration
by parts, use of the Bianchi identity and setting $k \alpha'=1$),
\be
S_{mod}=\frac{-1}{16 \pi \alpha'} \int d^4 x \sqrt{-g} e^{-\phi}
\biggl\{ c_0 R_{\mu \nu \lambda \sigma} R^{\mu \nu \lambda \sigma}+
 c_1 R_{\mu\nu} R^{\mu\nu}+c_2 R^2+c_3 (\nabla \phi)^4+ \nonumber
\ee
\be
+c_4 R^{\mu\nu}
\nabla_{\mu}\phi\nabla_{\nu}\phi+c_5 R(\nabla\phi )^2+c_6 R\Box\phi
+c_7 \Box\phi(\nabla\phi )^2+c_8 (\Box\phi )^2 \biggr\},
\label{gentens}
\ee
with
\beq
c_0 &=& 0 \nonumber \\
c_1 &=& -a_1 \nonumber \\
c_2 &=& \frac{a_1}{2}+a_3-2 b_1 \nonumber \\
c_3 &=& -a_2-2 a_4+2 b_2 \nonumber \\
c_4 &=& -a_1-a_2 \nonumber \\
c_5 &=& \frac{a_2}{2}-2 a_3 +a_4+2 b_1-2 b_2 \nonumber \\
c_6 &=& \frac{a_1}{2}+3 a_3+a_5-4 b_1-2 b_3 \nonumber \\
c_7 &=& \frac{3 a_2}{2}+3 a_4-2 a_5-4 b_2+2 b_3 \nonumber \\
c_8 &=& 3 a_5-4 b_3.
\eeq
We have included the $c_0$ term even though it is unchanged by
field redefinitions so we have a list of all independent covariant
tensors at this order.
As observed in \cite{cor,mag}, freely varying the $a$ and $b$ parameters
results in free variation of the non-zero $c$ parameters subject to the
single constraint
\be
c_2+c_3+c_7+c_8=c_5+c_6.
\label{redefconst}
\ee
We should remark that the reemergence of the $(\nabla \phi)^4$ term
does not contradict the calculation (\ref{tseytlin}) since its 
contribution to the S-matrix is offset by the other terms introduced.

These shifts are useful to exhibit forms of the action
which have equations of motion containing at most second derivatives.
Higher derivative equations of motion are difficult to handle, since
the extra initial conditions which must be imposed suggest that we've
allowed extra modes into the problem. Practically, these extra
modes lead to numerical instabilities and runaway solutions. 
This physical nature of this problem is lucidly discussed in \cite{wald} 
by analogy with the radiation reaction on an accelerating point charge,
which can produce similar runaway solutions. The equation of motion
becomes third order in derivatives, apparently introducing a new
degree of freedom, the choice of initial acceleration. 
However, the initial value
of the acceleration must be adjusted exactly to cancel an exponentially
runaway solution, which should be regarded as unphysical. So in 
spite of an apparent increase in the number of 
higher derivative initial conditions, the restriction to `physical'
behavior will eliminate them. In our case of higher derivative
corrections, it is practically impossible 
to find these special initial conditions with any exactitude, and
even if we could, numerical instability would render the solution
useless after a short time.
These problems can also be dealt with on a perturbative level by a
prescription called reduction of order, in which higher order
derivatives coming from the corrections are replaced by
forms obtained by differentiating the lowest order parts of the
equations of motion \cite{wald}. This leads to modified equations
of motion which formally differ from the original
by truncation of terms containing
higher powers of the perturbative expansion parameter $\alpha'$.
However, the modified equations, while of lower order, are often
extremely complicated and we will not explore this approach
further. Here we will simply explore only those forms of corrections
which do not introduce higher derivatives.

As we shall see later, making these shifts, while formally
preserving the action to $O(\alpha')$,
have a drastic effect of the behavior of solutions,
not only in the region of fixed points, causing fixed points to
move or even cease to exist, but making qualitative changes
in the perturbative
regime. Our knowledge of the form of the corrections at 
this point is not sufficient to answer the most basic 
questions about the behavior of the solutions. So we need
other information to constrain them further.

The existence of the inflationary (+) branch solutions
can be traced to a symmetry of the lowest order action
\cite{dual},
scale factor duality (SFD), which can be extended to a larger
symmetry $O(d,d)$ in the presence of the antisymmetric tensor
\cite{odd}. The origin of SFD lies in a canonical transformation
on the string world sheet and since the worldsheet fields will have this
symmetry, if we could untangle the fields relationships with the
redefined and renormalized fields in the effective action 
at a given order, we would see the symmetry realized in the 
corrections, perhaps in a non-trivial way \cite{odd}. 
So it is tempting to try to use this symmetry
to extract information about the unknown parts of the higher
order corrections. This subject has already been extensively
explored by Maggiore \cite{mag} and many of the following results were
originally reported there. Here we confirm them independently in 
a different setup and
make some additional observations.

Recall SFD in it's simplest (isotropic) form in 4 dimensions,
\beq
 {{\phi(t)}\rightarrow \phi(t) - {6\,\log (a(t))}} \nonumber 
\eeq
\beq
 {{a(t)}\rightarrow {{1\over {a(t)}}}}.
\label{naivesfd}
\eeq
We see that in terms of the variable
${\bar \phi}(t) = \phi(t) - 3\,\log (a(t))$, SFD takes the simple
form
\begin{eqnarray}
 {\bar \phi(t)} \rightarrow  {\bar \phi(t)} \\
 {H(t) \rightarrow -H(t)}.
\label{naivebar}
\end{eqnarray}
The equations of motion become
\beq
-3 H^2-\bar \rho+ \pbd^2 &=& 0 \label{rhoeq} \\
\bar \sigma-2 \dot H+2 H \pbd &=& 0 \label{sigeq} \\
\bar \lambda-3 H^2- \pbd^2+2 \pbdd &=& 0 \label{lameq} \\
3 \bar \sigma H+ (\bar \lambda - \bar \rho) \pbd + \dot {\bar \rho} &=& 0,
\label{coneq} 
\eeq
where $\bar \rho=e^{\phi} \rho$, $\bar \lambda=e^{\phi} \Delta_\phi\L_c$
and $\bar \sigma=e^{\phi} (p+\Delta_\phi\L_c)$. In addition to a slightly
simplified form, this version has the advantage that the terms of
each equation are uniformly even or odd under SFD.
So a source can easily be
inspected for SFD invariance, $\bar \rho$ and $\bar \lambda$ should
be even and $\bar \sigma$ should be odd. These conditions can be
guaranteed by showing that $\L_c$ can be written in a form that is
explicitly SFD even. Generally this will require integrations by parts
to eliminate total derivatives that don't have this property. For
example the lowest order part of the action can be explicitly
displayed in an SFD invariant form as
\beq
\Gamma =
\int dt e^{-\bar \phi} ( {{3 H^2}\over{n}}-{{\pbd^2}\over{n}} + \L_c) =
\int dt e^{-\bar \phi} \L_{eff}(H,\pbd,\dot H,\ldots).
\label{lowact}
\eeq

\subsection {Explicit Examples}

In this section we look at explicit examples of evolution with
various forms of constraints imposed on the corrections.
We remark that many of the interesting properties of these solutions
can be deduced without explicit numerical integrations. In the 'no higher
derivative' case the constraint equation (\ref{rhoeq}) is an algebraic
equation in the $(\pbd,H)$ phase plane. So solutions will be confined to 
flow on this curve. The location of the fixed
points can be found by intersecting this curve with the curve defined
by taking one of the other equations of motion and putting the
higher derivatives to zero. While this might seem to lead to
more equations than unknowns, as observed in \cite{gmv}, the
conservation equation (\ref{coneq}) is actually a linear relation
between the other three equations in a fixed point,
reducing the system to two equations
in two unknowns and allowing for the generic existence of fixed
points.

To map landmarks in the $(\pbd,H)$ plane, we solve the 
constraint equation (\ref{rhoeq}) for $\pbd$,
\beq
\pbd=\pm \sqrt{3 H^2+\bar \rho}.
\eeq
The sign choice here corresponds to our designations of (+) and 
($-$) branch respectively. So the vacuum solutions ($\bar \rho=0$) will
appear as straight lines $\pbd=\pm \sqrt{3} H$. Explicitly as
a function of time the expanding vacuum solutions are,
\beq
\pbd=-\frac{1}{t} ~~~ H=\mp \frac{1}{\sqrt{3} t}.
\label{expvac}
\eeq
The upper sign corresponds to the (+) solution ($t<0$) and the lower
to the ($-$) ($t>0$). 
The change from
one branch to another happens along the line $\pbd=0$. And, as discussed
in detail in \cite{bm1,bm2}, the line $\dot \phi=2 H$ or these
variables $\pbd=-H$, is where the `bounce' (the change from
expansion to contraction or vice versa) takes place in the 
Einstein frame (a conformally related frame in which the 
gravitational coupling is held constant). We are currently working
in the string frame in which the string scale $\lambda_s$ is a 
constant. It is the crossing of this line that is associated with
NEC violation. 

We will also find the solutions will sometimes encounter 
singularities at finite values of $\pbd$ and $H$. To see where these 
come from we solve the equations of motion (\ref{sigeq}) and (\ref{lameq}) 
for $\pbdd$ and $\dot H$ in terms of 
lower derivatives. Without higher order corrections this is trivial,
but when higher order corrections are added $\bar \sigma$ and 
$\bar \lambda$ can contain terms like $H^2 \pbdd$, $\pbd^2 \dot H$
etc. This means solving for $\pbdd$ and $\dot H$ can lead to expressions
containing denominators. Clearly if the solution approaches the 
curve corresponding to the vanishing of one of these denominators
the higher derivatives will go to infinity and the integration must 
be stopped. The curves also generally mark changes in the 
flow direction on the constraint curve. So we will also plot
curves indicating the vanishing of these denominators.

To begin with the simplest case we will also impose the requirement
that the action produces equations of motion having at most second
derivatives in the variables $a$ and $\phi$. This limits us to four
possible tensor structures in the correction.
\beq
 \L_c=k \alpha'
 ({A\,{(\nabla \phi )^4}}
 +
 {B\,{R^2}_{GB}}
 + 
 {C\,(R^{\mu
 \nu}-{1 \over 2} g^{\mu \nu} R) \,\nabla_{\mu} \phi \,\nabla_{\nu} \phi}
 + 
 {D\,\nabla^2(\phi)\,{{(\nabla \phi)}^2}}
 ) \label{lowderact}
\eeq
Where $R_{GB}=R^{\mu \nu \lambda \sigma} R_{\mu \nu \lambda \sigma} -
              4 R^{\mu \nu} R_{\mu \nu} + R^2$ is the Gauss-Bonnet
term. We can consider putting other constraints on this correction,
for example, the requirement that it contain the Riemann squared 
term of (\ref{tseytlin}) is
\beq
B={{1} \over {2}}. \label{refdefB}
\eeq
This is because the Riemann squared term is not altered by a field
redefinition.  
The requirement that the rest of the terms come from a field
redefinition of the form (\ref{redef}) from the basic correction
(\ref{tseytlin}) is,
\beq
A+B+C/2+D=0. \label{redefABCD}
\eeq

\bfig
\bigepstwo{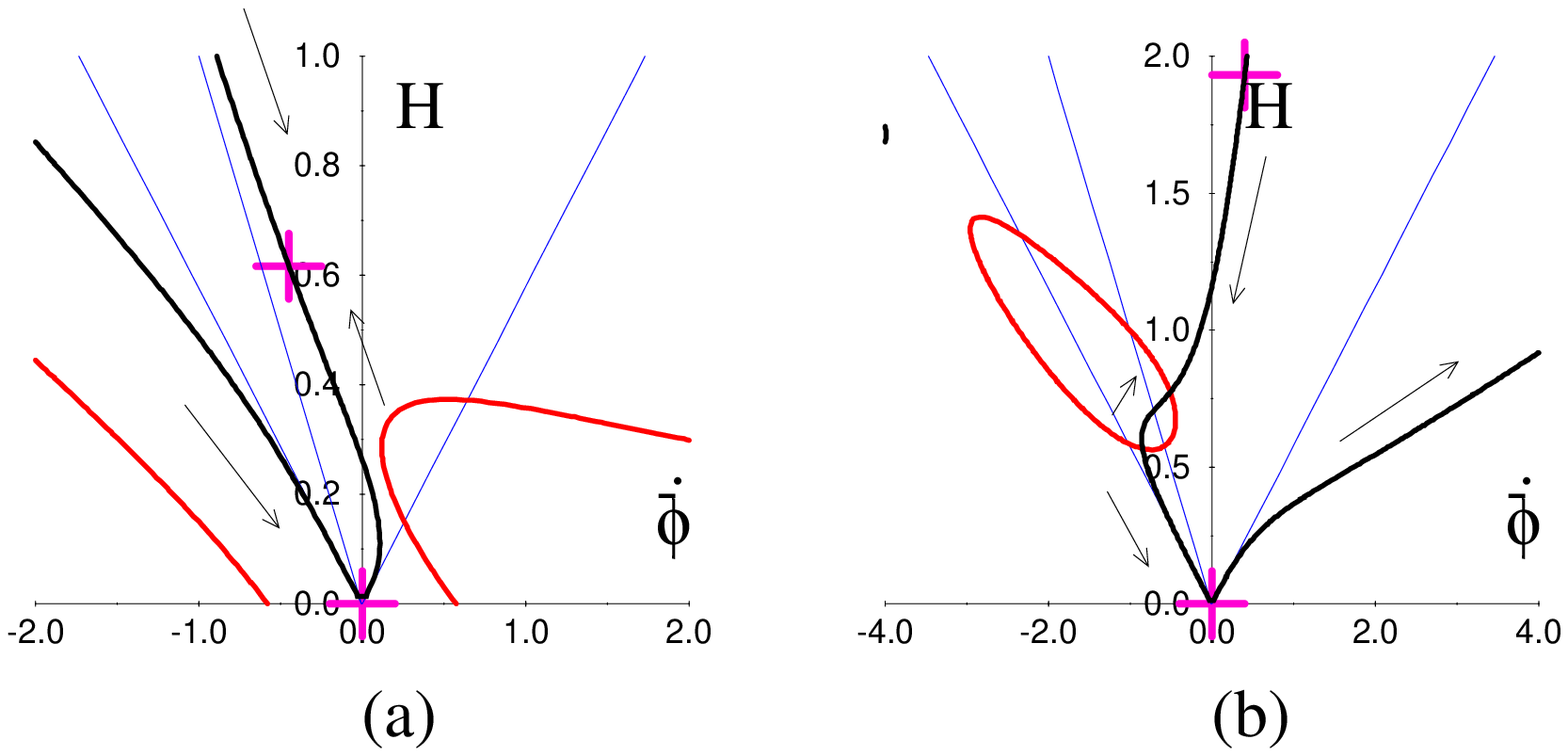}
\caption
{\small\it
Sample evolutions in the $(\pbd,H)$ phase space. The three blue lines
are from left to right, the $(-)$ branch vacuum, the Einstein frame
bounce line and the $(+)$ branch vacuum. The $H$ axis is the line of
branch change. The black line marked with arrows is the constraint
curve (\ref{n00eq}) with the arrows indicating flow direction. Fixed
points are marked with magenta crosses and the red curve indicates
lines of singularities where the denominator of the expression for
$\dot H$ and $\ddot \phi$ in terms of lower derivatives vanishes.
a) The effect of the $O(\alpha')$ corrections
in \cite{gmv}, $A=-1/2, B=1/2, C=0, D=0$ b) The effect of the
correction of \cite{odd} $A=-1/2, B=1/2, C=-2, D=1$ }
\label{f:fig0}
\efig

Before we look at the consequences of imposing SFD on our sources
we look at two examples of explicit evolutions of this type
in the $(\pbd,H)$ phase plane.
\figref{f:fig0}{a} shows the case $A=-1/2, B=1/2, C=0, D=0$, 
the case examined in \cite{gmv} and used as part of the
foundation for a model of graceful exit in \cite{bm2}. The (+) branch
vacuum flows into a fixed point located at $(-0.445,0.617)$ after
undergoing a branch change. 
\figref{f:fig0}{b} shows the case $A=-1/2, B=1/2,
C=-2, D=1$. This is the form of corrections proposed in \cite{odd}.
It has the remarkable property of being SFD (indeed $O(d,d)$) invariant,
but with the form of the duality (\ref{naivesfd}) modified by 
corrections of order $\alpha'$ and with terms of higher order
in $\alpha'$ truncated. So it does not show the symmetry of
(\ref{naivesfd}) and furthermore the (+) branch solution flows
away from the region of branch change and does not encounter a 
fixed point. 

The important thing to note here is that these corrections
are related to the correction (\ref{tseytlin}) by a `field redefinition'
(check (\ref{redefABCD})) yet show very different behavior. Not only
are they different in terms of fixed point behavior, very close to the
(+) branch vacuum the curves are turning in opposite directions.

To consider the effects of imposing (\ref{naivesfd}) on the correction,
we consider a correction with completely general $A, B, C, D$.
This adds to the quadratic terms in (\ref{lowact}) the correction,
\beq
 \L_{c}=\biggl\{{{3\,\left(
 27\,A 
 + 
 8\,B 
 + 
 9\,{
 C} 
 + 
 27\,{
 D} 
 \right) 
 \,{{H}^4}}\over
 {2\,{{n}^3}}}
  +
 {{\left(
 54\,A 
 + 
 4\,B 
 + 
 9\,{
 C} 
 + 
 45\,{
 D}
 \right) 
 \,{{H}^3}\,{\dot
 {\bar 
 \phi}}}\over
 {{{n}^3}}} \nonumber 
\eeq
\beq
 + 
 {{3\,\left( 
 18\,A 
 + 
 {
 C} 
 + 
 12\,{
 D} 
 \right) 
 \,{{H}^2}\,{{{\dot
 {\bar 
 \phi}}}^2}}\over
 {2\,{{n}^3}}}
 + 
 {{3\,\left( 
 2\,A 
 + 
 {
 D} 
 \right) 
 \,H\,{{{\dot
 {\bar 
 \phi}}}^3}}\over
 {{{n}^3}}}
 + 
 {{\left( 
 3\,A 
 + 
 {
 D} 
 \right) 
 \,{{{\dot 
 {\bar 
 \phi}}}^4}}\over
 {6\,{{n}^3}}}
 \biggr\}. \label{explowderact}
\eeq

The qualitative question of whether this action has 
solutions which turn towards
the branch change region ($\rho<0$) as in \figref{f:fig0}{a} 
or away ($\rho>0$) as in \figref{f:fig0}{b} is
easily answered by inserting the (+) branch vacuum solution
(\ref{expvac}) into (\ref{lowderact}) (since $\rho$ comes from the
variation by $n$, the $\rho$ contribution is proportional to 
the above form of the action). Numerically, we find the turning
direction is determined by the sign of,
\beq
\rho_0=83.5692 A+6.3094 B+11.1962 C+63.1769 D.
\label{rhozero}
\eeq
So if $\rho_0<0$ we expect the solution to turn towards the
branch change and conversely. Also, because the constraint
equation contains only terms of degree two and degree four when
we restrict ourself to only the $O(\alpha')$ corrections, we 
see the solution can have at most one nonzero intersection with 
every radial line through the origin. So once it turns one way
it will not turn back.

The requirement of SFD symmetry can now be imposed by forcing the
action to be SFD invariant. This is done by setting the
coefficients of the $H \pbd^3$ and $H^3 \pbd$ terms to zero, i.e.
\begin{eqnarray}
D&=&-2 A \nonumber \\
C&=&4 A-{4 \over 9} B.
\label{isosfd}
\end{eqnarray}

We then checked an observation \cite{mag}
that these corrections fail to satisfy
SFD in an anisotropic background, since allowing three different
Hubble constants in three directions $H_x$, $H_y$ and $H_z$ will create
many more SFD odd terms which must be set to zero. So we simultaneously
relaxed the 'no higher derivatives' conditions, which allows for the nine
different tensor structures not related by Bianchi identities
shown in (\ref{gentens}). The equations
become enormously more complicated, and since it is not clear which
integrations by parts should be performed to exhibit the action in
SFD invariant form (if indeed this is possible) we derived the
$\bar \rho$, $\bar \lambda$ and $\bar \sigma$ expressions and inspected
them for the correct SFD symmetry. We found a two parameter family of
corrections which did not break SFD invariance. In terms of the $c$'s
of (\ref{gentens}), the remaining seven coefficients of the 
SFD invariant corrections can be
parameterized in terms of the values of $c_3$ and $c_8$,
\beq
c_0 &=& 0 \nonumber \\
c_1 &=& 4 c_3 - c_8 \nonumber \\
c_2 &=& -c_3+c_8/2 \nonumber \\
c_4 &=& 4 c_3-c_8 \nonumber \\
c_5 &=& -2 c_3 \nonumber \\
c_6 &=& c_8 \nonumber \\
c_7 &=& -2 c_3-c_8/2.
\eeq
The condition $c_0=0$ makes this incompatible with the calculated
result (\ref{tseytlin}).
Further, the only combination of the curvature squared terms not giving
rise to higher derivative equations of motion is the Gauss-Bonnet 
combination, which in turn requires $4 c_0=4 c_2=-c_1$, so there are
no nontrivial members of this family without higher derivative equations
of motion.

With these cautions, we still might hope that the family of corrections
given by (\ref{isosfd}) might give some clue as to the nature of the
correct corrections, and they might indicate
that the solutions coming out of the
(+) branch vacuum tend to wind up in fixed points. Numerical
examinations showed this was not the case.
We present a representative family of such evolutions in
\figref{f:fig1}{}. Notice the enhanced symmetry over 
\figref{f:fig0}{}. In \figref{f:fig1}{a,d} we see an attractive
and repulsive fixed point on the $\pbd$ axis at $\pbd$ negative
and positive respectively. However they lie on a portion of 
the constraint curve
disconnected from the vacuum part. In \figref{f:fig1}{b,c} 
the vacuum part of the constraint forms a closed loop, 
and we could
hope it performs the graceful exit on its own. But this loop
is cut by the singularity curve where the integration must be halted.

\bfig
\bigepsfour{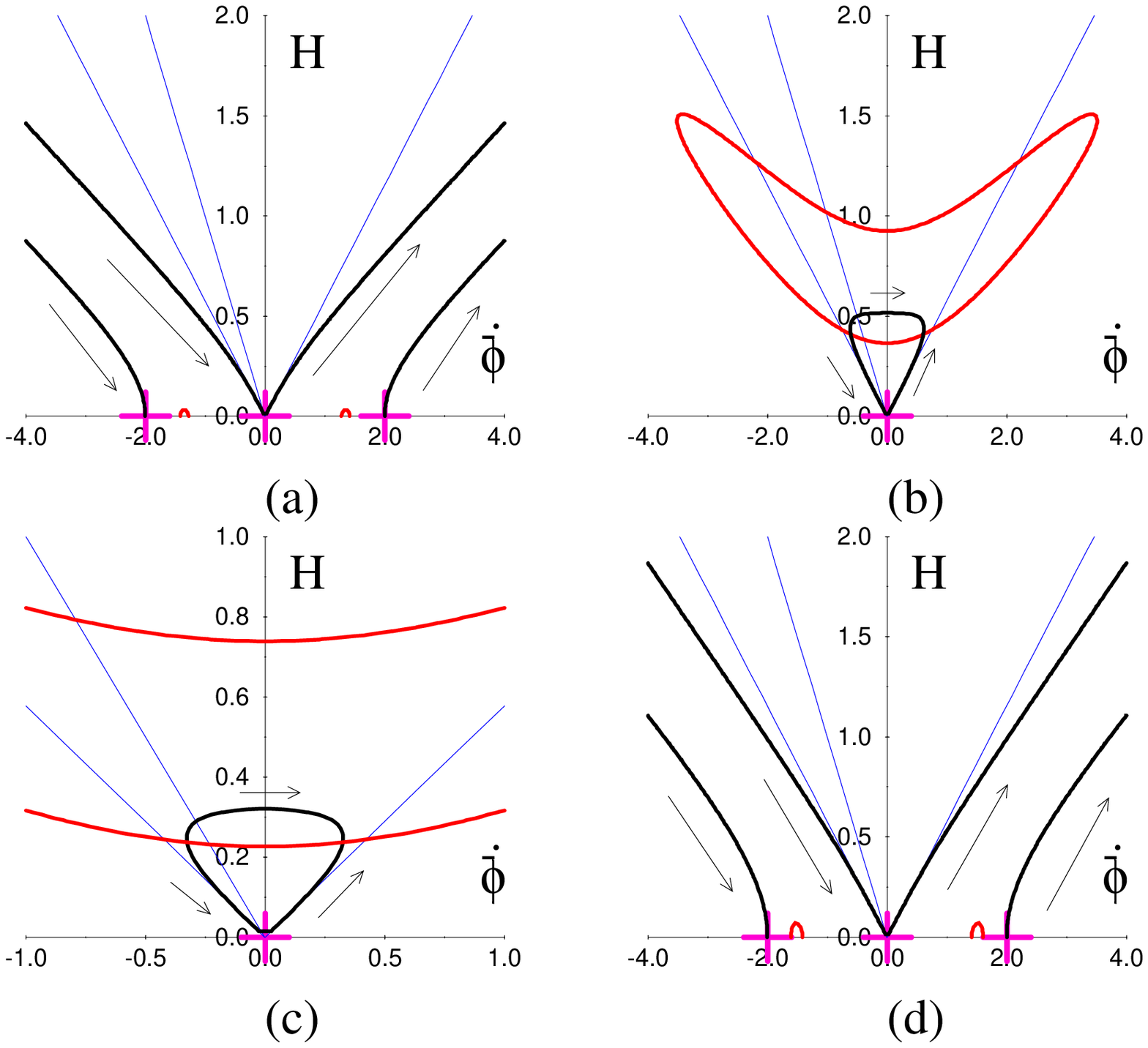}
\caption
{\small\it
Sample evolutions of corrections constrained by (\ref{isosfd}).
Features plotted are as explain in \figref{f:fig0}{}.
a) A=1/2, B=1/2. b) A=-1/2, B=1/2. c) A=-1/2, B=-1/2. d) A=1/2, B=-1/2.
}
\label{f:fig1}
\efig

\subsection {Naive SFD is a Bad Thing}

One conclusion might be that this shows that the order $\alpha'$
corrections are insufficient to describe the behavior and we
need to know the higher order corrections as well. This may
be, and indeed the fixed points generally occur at regions of the phase
space where the order $\alpha'$ corrections are on the same order
as the lowest order terms. But there is an even more fundamental
obstruction. We will now show that SFD in its naive form (\ref{naivesfd}) 
is generically incompatible with good fixed point behavior.

First of all we note another fundamental symmetry of our equations
of motion, time reversal invariance
\begin{eqnarray}
 {t \rightarrow  -t} \nonumber \\
 {\pbd(t)} \rightarrow  -{\pbd(-t)} \nonumber \\
 {H(t) \rightarrow  -H(-t)}.
\end{eqnarray}
This tells us that if we have a solution curve in the ($\pbd,H$) plane,
we can reflect it through the origin and reverse the time flow to
obtain another solution. 
Putting this together with SFD (\ref{naivebar}), which says we can
also reflect solutions across the $\pbd$ axis we see solutions
can also be reflected across the $H$ axis and time reversed. This 
in turn tells us the fixed points also occur in pairs reflected
across the $H$ axis with one repulsive and one attractive as a 
result of the time reversal.

Now we repeat an observation \cite{gmv} that the lowest order
action (\ref{lowact}) is independent of $\beta=\log(a(t))$,
and depends only on its derivatives,
all explicit $\beta$ dependence  having been absorbed
into $\bar \phi$. We presume this independence will persist in the
higher order corrections. This allows us to drop the first term of the
variation,
\def\empb{ e^{-\bar \phi}}
\def\epb{ e^{\bar \phi}}
\def\ddtone{{{d} \over {dt}}}
\def\ddttwo{{{d^2} \over {dt^2}}}
\beq
\delta \Gamma=\int dt \empb \left[
\partby{\L_{eff}}{\beta}{\delta \beta}+\partby{\L_{eff}}{\dot \beta}{
\delta \dot \beta}+\partby{\L_{eff}}{\ddot \beta}{\delta \ddot \beta}+\ldots
\right] \label{qcond},
\eeq
so after an integration by parts,
\beq
\delta \Gamma=\int dt \left[ -\ddtone (\empb \partby{\L_{eff}}{\dot \beta})
+ \ddttwo (\empb \partby{\L_{eff}}{\ddot \beta}) + \ldots \right] 
\delta \beta=0.
\label{varbeta}
\eeq
Since the quantity in brackets is just proportional to the $\beta$ equation
of motion (\ref{sigeq}) with an overall factor of $\empb$ and
it is clearly a total derivative, we can integrate it to get
a constant of motion \cite{gmv},
\beq
Q=\int dt \empb (\bar \sigma-2 \dot H+2 H \pbd)= \empb (\bar \Sigma-2 H),
\label{conofmo}
\eeq
where
\beq
\bar \Sigma(H,\pbd,\dot H.\ldots)=\epb \int dt \empb {\bar \sigma}
\eeq
is a function beginning at third degree in derivatives with the
$O(\alpha')$ corrections. We will find the this conserved constant
will allow us to make some statements about the location of fixed
points.

Putting $Q=0$ in (\ref{conofmo}) leads to another constraint type
equation for the solution, $\bar \Sigma-2 H=0$. This clearly conflicts
at lowest order with the constraint (\ref{rhoeq}), so this is not
a possibility for solutions originating near the vacuum. Considering
a solution approaching a fixed point, we have $\bar \Sigma - 2 H \rightarrow
constant$. If $\pbd>0$ then clearly the decreasing exponential
in (\ref{conofmo}) will
force $Q$ to be zero, so the only remaining possibility is $\pbd \le 0$
and $\bar \Sigma-2 H=0$ in the fixed point. So the diverging exponential
is cancelled by the decrease in $\bar \Sigma-2 H$, but now it would
appear we have another algebraic condition
to impose on a fixed point, raising
the question again of whether fine tuning is necessary to get a fixed
point. But we can show this is not the case.
The content of the equation (\ref{sigeq}) is
equivalent to
\beq
I_1=\epb \left[ \ddtone{ ( \empb \partby{\L_{eff}}{H} ) } -
   \ddttwo{ ( \empb \partby{\L_{eff}}{\dot H} ) }+\ldots \right] = 0.
\label{acteom}
\eeq
The integrated condition $\bar \Sigma - 2 H = 0$ which we must enforce
at fixed points is proportional to
\beq
I_2=\epb \left[ ( \empb \partby{\L_{eff}}{H} ) -
   \ddtone{ ( \empb \partby{\L_{eff}}{\dot H} ) }+\ldots \right] = 0.
\label{intacteom}
\eeq
In a fixed point, the time derivatives of the partial derivatives
of $L_{eff}$ vanish, since they are functions of $H$, $\pbd$ which
are becoming constant and derivatives higher than the first will vanish
in a fixed point. So the only non-vanishing parts will come from the
time derivatives of the $\empb$, so $I_1= -\pbd I_2$. So at least
in the case where $\pbd \ne 0$ in the fixed point, the vanishing of
the equation of motion $I_1$ implies the vanishing of $\bar \Sigma
- 2 H$.

   Putting these arguments together shows that fixed points at $\pbd<0$
are attractive in the sense that generic ($Q \ne 0$) solutions will
flow into them. Conversely, solutions at $\pbd>0$ are repulsive in the
sense that solutions flow out of them. We have not found an argument
classifying the behavior at fixed points at $\pbd=0$. In addition, we
have the possible existence of $Q=0$ solutions which evade these
constraints. We shall have more to say about these when we confront
one in section 3.

\bfig
\bigepsfour{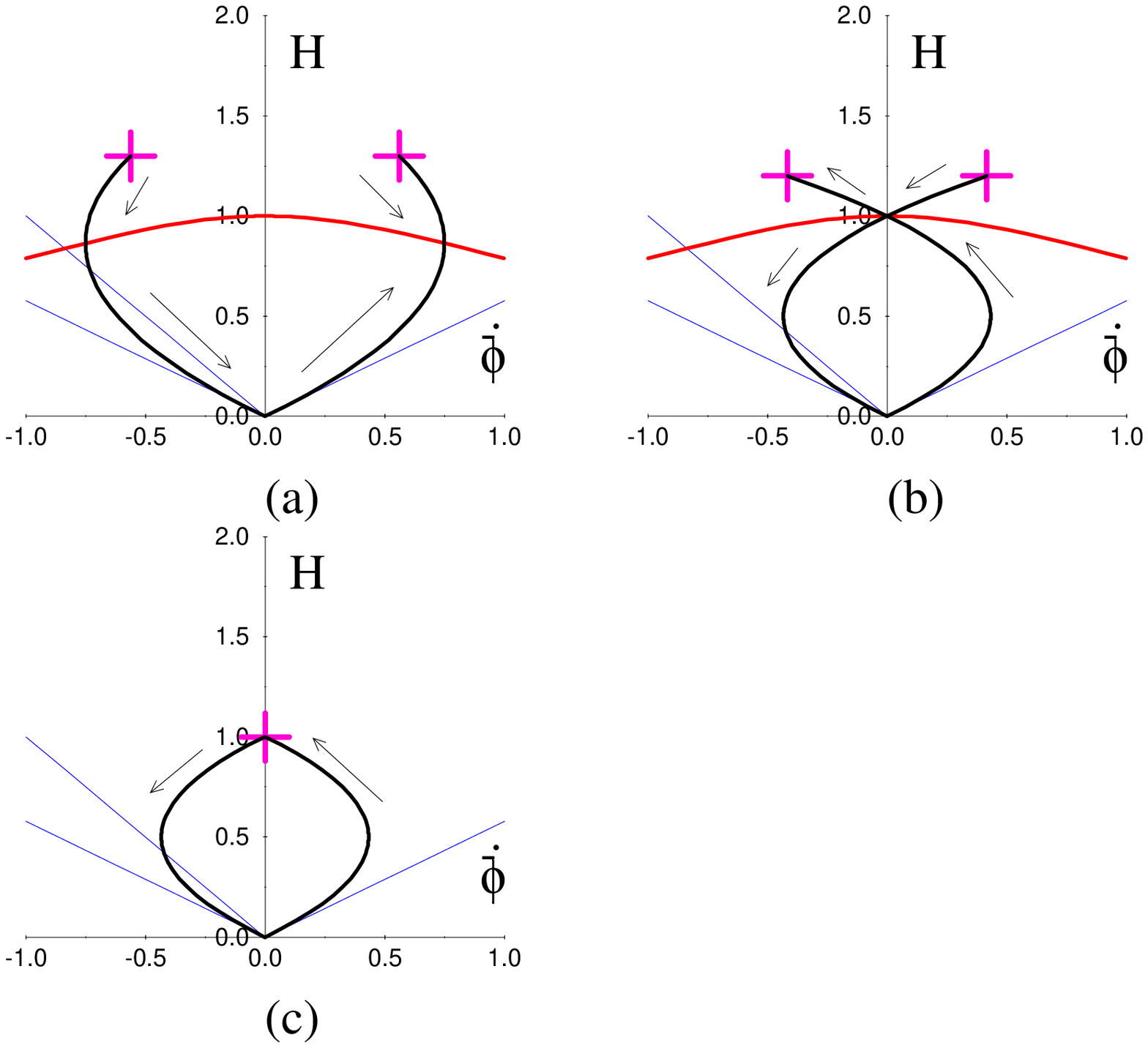}
\caption
{\small\it
Possible cases of evolution to and from fixed points in the case
of a self dual action.}
\label{f:fig2}
\efig

Now we are ready to discuss the possibility of a solution coming
out of the $(+)$ branch vacuum and ending in a first fixed point 
encountered on the constraint curve for an
SFD invariant action. Given the symmetries of SFD invariant actions
we can draw three possible pictures of such a solution and 
it's SFD/time reversed partner, a solution
flowing out of a fixed point and 
into the ($-$) branch vacuum \figref{f:fig2}{}.
In \figref{f:fig2}{a} the fixed point is at $\pbd \ge 0$.
Therefore, as we have argued, this fixed point is repulsive,  
and a solution will flow out of it rather than in.
Given that a solution is also flowing out of the (+) branch
vacuum, they must meet at some intermediate point. 
This point will not be a regular fixed point since we presume the
cross marks the first fixed point.
The only other possibility is that it
is a singularity of the type discussed in \cite{bm2,mag},
which is manifested in the equations of motion
by a zero in the denominator of the expression for the highest
order derivatives, $\dot H$ and $\pbdd$
in terms of lower order derivatives, and which we have indicated
by drawing such a curve. So this case leads to 
singular behavior rather than fixed point behavior.

In the second case, \figref{f:fig2}{b} the problem is a little 
more subtle. Here we have put the first fixed point at $\pbd<0$, 
so it is attractive, but something
peculiar is clearly going on where the two solutions cross.
We have two solutions leaving the same ($\pbd,H$) point,
when we would expect that the value of $\pbd$ and $H$ would
uniquely specify initial conditions. This is because first, 
we have presumed that equations are at most second order in
derivatives. Second, the variables $a(t)$ and $\phi(t)$ do not
appear explicitly in the equations of motion.
In fact, this
situation is the same as the singularity in the first case.
The expressions for $\dot H$ and $\pbdd$ take the
indeterminate form $0/0$ at this point. So the curve corresponding to the
vanishing of the denominators of these expressions also passes
through this point, and higher derivatives go to infinity
in the neighborhood of such points.
Numerical simulations are wildly unstable
passing through there, and we regard it as physically unstable
as well. 

It might be objected that this looks like the behavior at the origin.
This is indeed possible if this crossing point is also a $\pbd=0$
fixed point, like the origin. In this case the solutions don't
actually cross, but just asymptotically approach this point.
This is the boundary behavior of both of the previous cases
when the fixed points are allowed to approach each other. This
is an interesting place for a fixed point, it is mapped into itself
by a combination of SFD and time reversal and so is both a $(+)$ and
$(-)$ branch solution simultaneously.

We now discuss briefly the possiblity of placing the fixed
point at $\pbd=0$, \figref{f:fig2}{c}, 
the only possibility that would allow us
to retain both simple SFD (\ref{naivesfd}) and good fixed point
behavior. To do this we need to examine the equations of motion
with all derivatives higher than the first set to zero, and with
$\pbd=0$. A little thought will show that the only terms in the
action that can contribute to these reduced eoms are
\beq
\int dt e^{-\bar \phi} ({{3 H^2}\over{n}} + c_1 {{H^4}\over{n^3}}
  + c_2 {{H^6}\over{n^5}}+\ldots ),
\eeq
where the dependence on $n(t)$, the lapse factor, is dictated by
time reparameterization invariance.
Performing the variation with respect to $n$, as in
(\ref{rhoeq}), and with respect $\bar \phi$, as in (\ref{lameq}),
and setting $n=1$, we get
the following,
\beq
3 H^2+3 c_1 H^4 +5 c_2 H^6+\ldots &=& 0 \\
3 H^2+c_1 H^4 + c_2 H^6+\ldots &=& 0.
\label{zerofix}
\eeq
So we see there are no non-trivial $\pbd=0$ fixed points at order $\alpha'$
(allowing us at most the power $H^4$).
At $O(\alpha'^2)$ it would seem to be possible. But we
have not classified covariant tensor structures at this order and
have no results comparable to (\ref{tseytlin}). Even so we have
tried to introduce SFD invariant corrections of the form expected
at $O(\alpha'^2)$ and while we can position a fixed point at
$\pbd=0$ we have found no cases where it well behaved and
connected to the vacuum. This should be expected, since
this case be looked at as a limiting case of the two
previous cases, where the action is manipulated to allow
the two fixed point to approach each other. Since the
previous cases exhibit pathologies, they should probably
be expected to persist in the limit, but we don't regard
this as a rigorous argument.

To summarize, we have shown that good fixed point behavior and
SFD with eoms containing derivatives at most of order two
require fixed points at $\pbd=0$, which in
turn seems to be difficult to achieve. 

On the other hand, it is not difficult to
find good fixed point behavior (as in \cite{gmv}) if SFD is broken.
So to sum up, perversely we
have the situation where imposing a string theoretical notion
(SFD) on possible classical corrections seems to create
difficulties for achieving another string theory notion, that
finite size string effects will saturate curvature growth.

There are several different conclusions to be drawn. One is simply
to accept these conclusions and retain faith in both the
simplest form of SFD and curvature saturation. Perhaps if
and when we can determine the correction
to all orders it will exactly solve (\ref{zerofix})
allowing a $\pbd=0$ fixed point and shed pathologies.
A second is the suggestion
of \cite{mag}, that we must consider the contribution of the
massive string modes, which may help to saturate curvature
growth.
A third direction lies in a modification
of the form of duality. As discussed in \cite{dual}, the source of
SFD is a classical symmetry involving the exchange of winding and
momentum modes on the string world sheet, so at this level the dilaton
does not participate in SFD at all, since the dilaton only enters
the theory at the quantum level. The non-trivial transformation of
the dilaton comes at the level of the effective action, where we are
working with `renormalized' fields, which mix the dilaton with the
scale factor. Working to higher orders in the corrections, one should
expect additional renormalizations and hence corrections to the form of
SFD. This expectation was exploited in \cite{odd} to actually fix
the form of the action and correction at $O(\alpha')$, based on correct
order by order cancellation of non-$O(d,d)$ invariant terms. 
As we have seen, this form of correction does not lead
to fixed point behavior (\figref{f:fig0}{b}). But as we will see
in the next section, it is close to the region of parameter space
that admits fixed points without SFD. Further, considering
(\ref{lameq}) we see that the source $\bar \lambda$ has to be at least
of the same order as $H^2$ or $\pbd^2$, so the corrections cannot be 
small in a fixed point. This suggests that rather than simply 
conclude SFD works against fixed point behavior,
we will require more knowledge of
higher order corrections and/or of the form scale 
factor duality takes in the
higher order effective action. 

\subsection{The Distribution of Fixed Points}

Since at this level SFD does not lead to interesting
conclusions regarding the fixed point behavior, we return
to the general form of corrections 
(\ref{lowderact}) and simply ask the question, what region
of the ($A,B,C,D$) parameter space for the corrections
leads to good fixed point behavior and where are the 
fixed points located? Since this is a four dimensional 
space we also impose the conditions
(\ref{refdefB}) and (\ref{redefABCD}), reducing us to 
a two parameter space ($A,D$). 
To find these fixed points one should
be careful not only to check algebraically that the fixed points
exist but that they are reachable from solutions
beginning near the (+) branch vacuum. To do this we run
numerical integrations from initial conditions near the
(+) branch vacuum and examine the solutions for fixed 
point behavior at late time. In \figref{f:fig3}{a} we have
placed a dot on a grid where an ($A,D$) selection leaves to 
good fixed point behavior. We have also marked the line corresponding
to the condition (\ref{rhozero}). As expected, there is no good fixed
point behavior to the right of this line, since the solutions initially
veer from the vacuum away from branch change, hence towards
$\pbd>0$ where we find only unstable fixed points. But good fixed
point behavior almost saturates the region to the left.

We have also marked with crosses the corrections corresponding to
the $\alpha'$ corrections used to produce fixed point behavior in
\cite{gmv,bm1,bm2} inside of the region of good fixed parameters,
and also the parameters corresponding to the corrections of \cite{odd},
whose evolution is plotted in \figref{f:fig0}{b}, which lies outside
of this region. This form of the corrections is perhaps the best 
motivated form of the corrections, coming from an exact but truncated
form of $O(\alpha')$ modified duality. It is discouraging that it
is outside the region of good fixed point behavior, but encouraging to
note that it is not far away, suggesting higher order corrections 
could easily modify it's behavior.

In \figref{f:fig3}{b} we plot the locations of the resulting fixed 
points in the ($\pbd,H$) plane and the location of the fixed point
used in \cite{gmv,bm1,bm2}.
We find that the good fixed points are all located in a wedge bounded
by $\pbd \le 0$ and $H \ge -\pbd$. The first boundary
is easy to understand, since we have just shown that the stable
fixed points must be located at $\pbd \le 0$. The second is harder
to understand. But we do observe that the line $H=-\pbd$ is
just the line where the scale factor undergoes a bounce
in the Einstein frame, and producing this bounce requires
the sources to violate the Null Energy Condition ($\rho+p \ge
0$) \cite{bm1,bm2}. As these sources represent classical string corrections
which are not expected to violate NEC,
it is possible these constraints contain part of that
condition.

\bfig
\bigepstwo{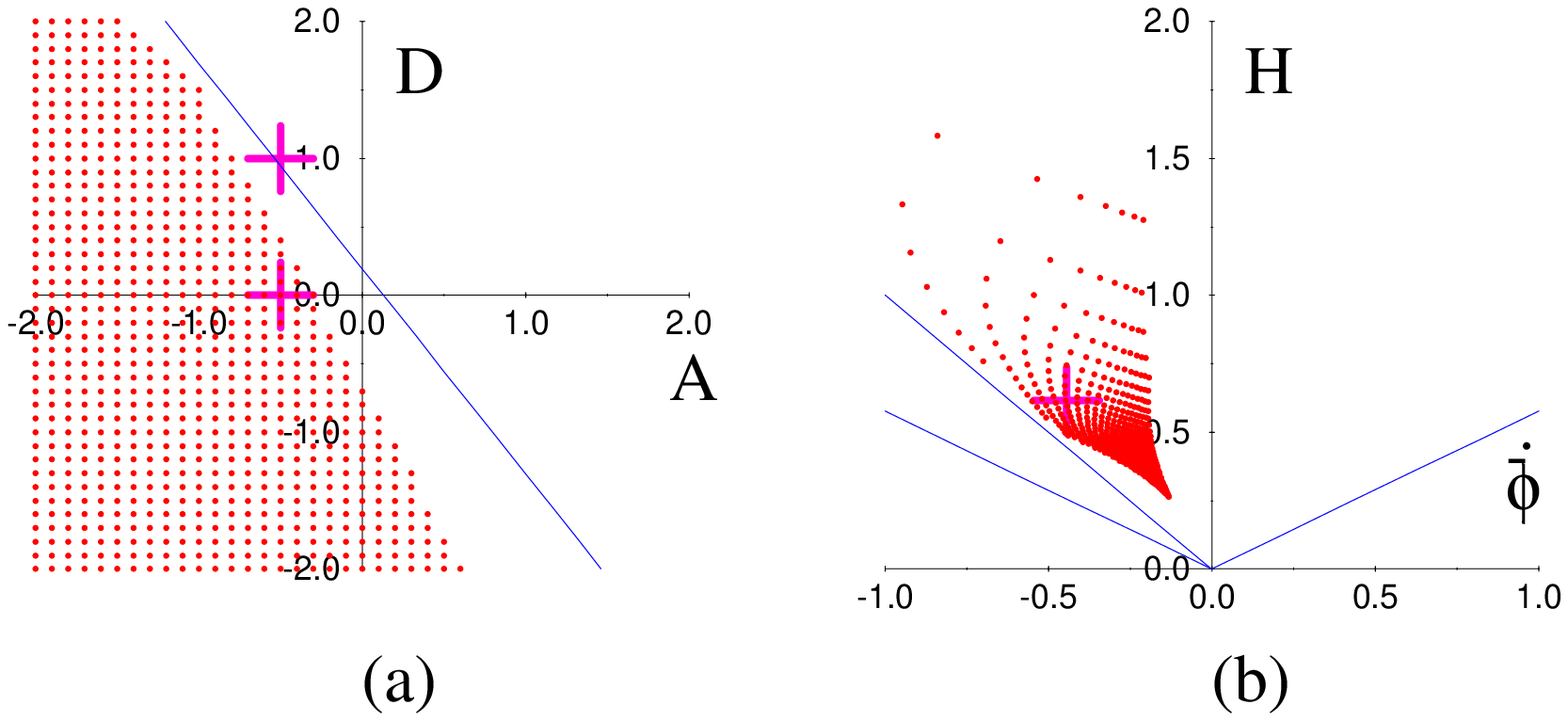}
\caption
{\small\it
a) The distribution of coefficients leading to good fixed point
behavior in the ($A,D$) plane. $C$ and $D$ are chosen to fit the
constraints (\ref{refdefB}) and (\ref{redefABCD}).
The line plotted is the condition of (\ref{rhozero}), to the
right of this line solutions turn away from branch change and
are not expected to lead to good fixed point behavior.
The crosses marks the location of the
corrections figured in \figref{f:fig0}{}. b) The location in the
($\pbd,H$) plane of the
resulting good fixed points. The cross marks the location of the
fixed point shown in \figref{f:fig0}{a}. }
\label{f:fig3}
\efig

\renewcommand{\theequation}{3.\arabic{equation}}
\setcounter{equation}{0}
\section {Conformal Field Theories}

We have seen that investigations of the behavior of solutions including
only $O(\alpha')$ corrections leads to, at best, ambiguous
conclusions about their fixed point behavior. 
Since the classical equations of motion for the background fields of the string
are derived from the requirement that they preserve conformal invariance
on the string worldsheet \cite{cor1}, 
directly constructing a conformally invariant
background would give a solution to all orders in $\alpha'$, even
without knowledge of the form of the corrections. While there is a large
literature of exact cosmological solutions
\cite{cosmocft} (coming, for example, from gauged WZW models), all exhibit
either extreme anisotropy or are supported by other fields in addition
to the graviton and the dilaton. So they are not particularly relevent to
this scenario. We will take a more naive and non-rigorous approach to 
constructing a background which we hope will at least have properties
in common with a conformally invariant background.

Recalling the classical action for the bosonic string,
\beq
S=\frac{1}{4 \pi \alpha'} \int {d \tau} {d \sigma} \sqrt{\gamma} \left[
\gamma^{\alpha \beta} G_{\mu \nu}(X) \partial_\alpha X^\mu \partial_\beta X^\nu
+ R^{(2)}(X) \Phi(X) \right],
\label{stringact}
\eeq 
where $\tau$ and $\sigma$ are coordinates and $\gamma$ is the metric
on the string worldsheet, $X^\mu$ and $G^{\mu \nu}$ are the spacetime
coordinates and metric, $R^{(2)}$ the worldsheet curvature and $\Phi$
is the dilaton.  
With $G^{\mu \nu}=\eta^{\mu \nu}=(-,+,+,\ldots)$ 
this leads to a conformally invariant
theory in critical dimensions where the reparameterization ghosts cancel
the contribution to the central charge of the bosonic fields $X^\mu$.
In the following we ignore issues related to the central charge of
the model, expecting it can also be cancelled by the addition of other
sources that are `inert', in the sense of not affecting the other 
conclusions we draw. We introduce a nontrivial background in the
above by assuming
\beq
\Phi(X)=P X^{0}=P t,
\eeq
where P is a constant, giving a dilaton varying linearly in time. 
We also add to the flat space action
((\ref{stringact}) with $G^{\mu \nu}=\eta^{\mu \nu}$) the term,
\beq
{\cal O}_K(z,\bar z)=\sum\limits_{i > 0} \gamma^{\alpha \beta} e^{2 K X^{0}} 
\partial_\alpha X^i(z,\bar z) \partial_\beta X^i(z,\bar z).
\label{addterm}
\eeq 
This leads to an action with a total background metric of FRW form with
$a(t)=\sqrt{1+e^{2 K t}}$, interpolating between 
flat space and a de-Sitter form
like our expected fixed point solutions. But we will need to insure that
the addition of (\ref{addterm}) to the action hasn't spoiled conformal
invariance.

A first step in this direction is to check that quantum effects do not
change the classical scaling dimension of (\ref{addterm}). 
A standard framework for doing this falls under the name of conformal
perturbation theory (see, for example, \cite{cft}). 
The energy momentum tensor for the flat space action (\ref{stringact})
with the linear dilaton ansatz is,
\beq
T_{z z}=-\frac{1}{2} \left( \partial_z X^0 \partial_z X^0-
\sum\limits_{i>0} \partial_z X^i \partial_z X^i+
P \, \partial^{2}_{z} X^0 \right),
\eeq
where we have used conformal invariance to put the world-sheet metric 
into the conformal gauge ($\gamma_{z \bar z}=\gamma_{\bar z z}=1,
\gamma_{z z}=\gamma_{\bar z \bar z}=0$). 
There are also exactly parallel formulae 
for the anti-holomorphic parts ($\bar z$), which decouple from the
holomorphic parts.

The requirement that ${\cal O}_K$ transform as a conformal tensor is
just,
\beq
{\cal O}_K(z,\bar z) \rightarrow \left( \partby{f}{z} \right)^h
\left( \partby{\bar f}{\bar z} \right)^{\bar h} 
{\cal O}_K(f(z),\bar f(\bar z)).
\label{conscaling}
\eeq
Since we require the action to be an invariant, we want $h=\bar h=1$
to offset the scaling of the integration measure ${d z}{d \bar z}$.
Since $T_{zz}$ is the generator of conformal transformations, it can
be shown that (\ref{conscaling}) requires the following singularity
structure in the operator product expansion,
\beq
T_{zz}(z) {\cal O}_K(w,\bar w)=\frac{h}{(z-w)^2} {\cal O}_K(w,\bar w)
+\frac{1}{(z-w)} \partial_w {\cal O}_K(w,\bar w)+{\rm non-singular}.
\label{opprod}
\eeq
As usual this 
can be related to a normal ordered product by contracting operators
as specified by Wick's theorem and using the following `mnemonic' for 
the propagators,
\beq
<X^\mu (z) X^\nu (w)>=\eta^{\mu \nu} \log(z-w).
\eeq
This is a mnemonic in the sense that it correctly represents the 
short distance behavior of the propagator and the operators are
to be regarded as normal ordered in the sense that we do not include
divergent contractions of operators with the same argument.

So the expression corresponding to the holomorphic part of the 
$i^{th}$ component of the left hand side of (\ref{opprod})
becomes,
\beq
-\frac{1}{2} \left( \partial_z X^0(z) \partial_z X^0(z)-
\partial_z X^i(z) \partial_z X^i(z)+
P \, \partial^{2}_{z} X^0(z) \right)
e^{2 K X^{0}(w)} 
\partial_w X^i(w)
\eeq 
The contractions are easily carried out because of the simple behavior 
of the exponential under contractions. The result is,
\beq
\frac{e^{2 K X^0(w)}}{(z-w)^2} \left( \partial_z X^i(z) - K (P+2 K)
\partial_w X^i(w) \right) + \frac{e^{2 K X^0(w)}}{(z-w)} 
2 K \partial_z X^0(z) \partial_w X^i(w)
\eeq
If we insert the Taylor expansion, 
$\partial_z X^\mu (z)=\partial_w X^\mu (w)+
  (z-w) \partial^{2}_{w} X^\mu (w) \ldots$, we recognize this as,
\beq
\frac{(1-K (P+2 K)) {\cal O}_K(w)}{(z-w)^2}+
\frac{\partial_w {\cal O}_K(w)}{(z-w)}
+{\rm non-singular}.
\eeq
Comparing this with (\ref{opprod}) we identify ${\cal O}_K$ as a conformal
tensor of dimension $1-K (P+2 K)$, so we can nontrivially satisfy the
requirements of conformal invariance by setting $P= - 2 K$.

Next to make contact with $\phi$, the dilaton in our effective 
action, we compare our equations of motion with those of \cite{cor1}.
We conclude that $2 \Phi=\phi$. 
Reading off the space-time metric and dilaton, 
\beq
\phi(t)=2 P t \nonumber \\
P=-2 K \nonumber \\
a(t)=\sqrt{1+e^{2 K t}} \nonumber \\
H(t)=\frac{K}{1+e^{-2 K t}} \nonumber \\
\pbd(t)=-4 K-3 H(t)
\label{cftsol}
\eeq

\bfig
\bigepstwo{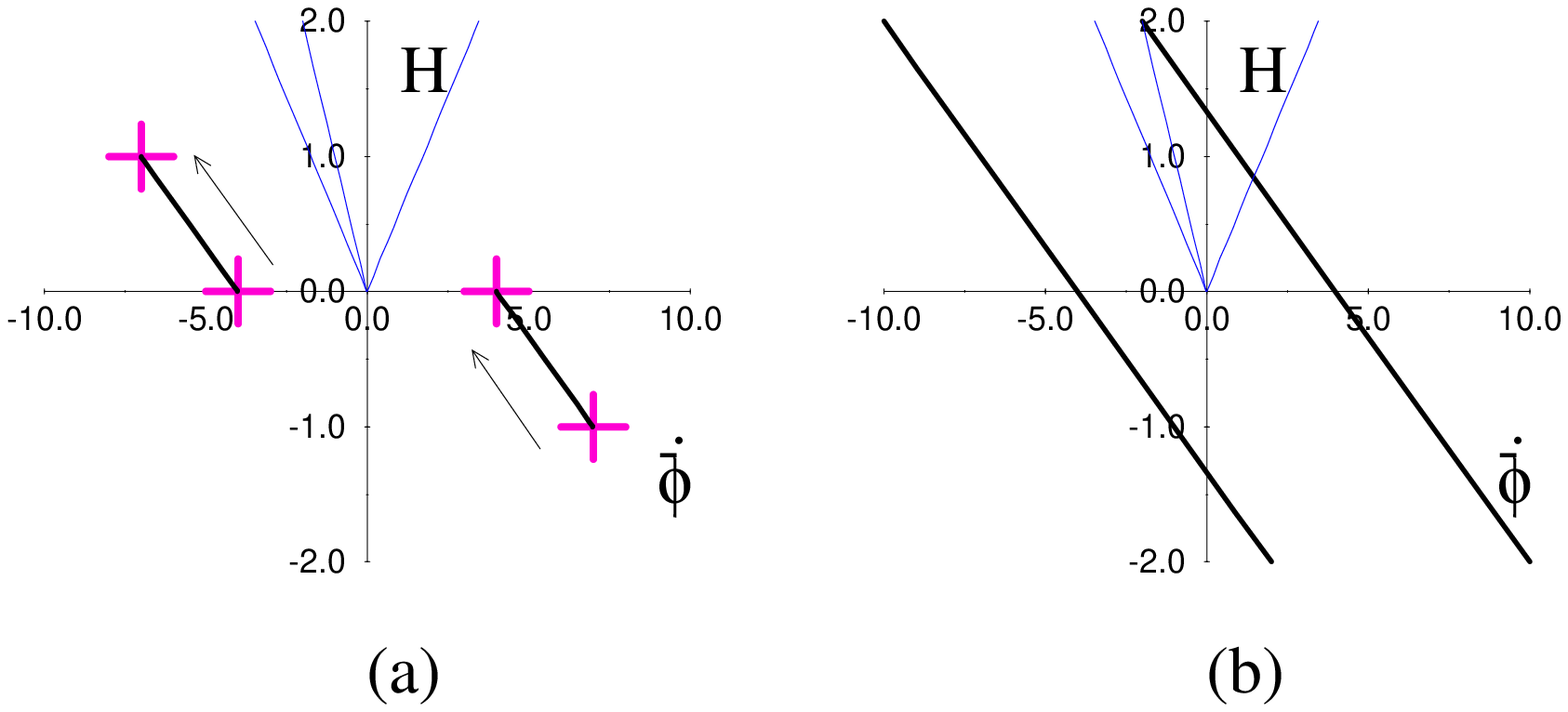}
\caption
{\small\it
a) The CFT trajectory, K=1, on the left and its time reversed partner
on the right, with beginning and ending fixed points marked. b) The CFT 
trajectory extended to include all of the fixed points allowed by the
equations of motion from an effective action. }
\label{f:fig4}
\efig
 
A sample of such an evolution is given in \figref{f:fig4}{a}. It
is unusual in light of our previous results. Its fixed points sit
beyond the line $H=-\pbd$ where the Einstein frame bounce requires
NEC violation, so we perhaps should not expect solutions
originating near the vacuum to flow into this fixed point. Secondly,
it is flowing out of a fixed point at $\pbd<0$, the opposite of the
expected behavior, as we discussed in section 2.3. 
So we expect that if it is in fact a solution to
an effective action, we will find it is a $Q=0$ solution.

To try to understand the implications of such a solution on the form
of the effective action we construct the most general effective action
not involving higher derivatives and attempt to fix the coefficients
by demanding this solution solve the eoms. We took the action to be of
the form,
\beq
\Gamma =
\int dt e^{-\bar \phi} \L(H(t),\pbd(t),n(t)) \nonumber \\
\L=\sum\limits_{n,i} c_{n,i} H(t)^i \pbd(t)^{n-i} / n(t)^{n-1},
\label{ansatz}
\eeq
where the sum is over even values of n.
As before, the power of the lapse factor $n(t)$ is dictated by
time reparametrization invariance. This is the most general action
that can be produced by covariant tensor corrections that do not
introduce higher derivatives. We can explicitly display the equations
of motion in terms of $\L$.
\beq
\partby{\L}{n}=0 \label{partn} \\
\parttbyo{\L}{H} \dot H+\parttbyt{\L}{H}{\pbd} \pbdd-\pbd
  \partby{\L}{H}=0 \label{parta} \\
\parttbyt{\L}{H}{\pbd} \dot H+\parttbyo{\L}{\pbd} \pbdd-
\pbd \partby{\L}{\pbd}+\L=0 \label{partphi}.
\eeq
We then fix $c_{2,2}=3$, $c_{2,1}=0$ and $c_{2,0}=-1$ reflecting
our knowledge of the lowest order action and add a finite
number of terms with even $n$, insert (\ref{cftsol}) into the
resulting equations of motion and attempt to solve the resulting
linear system for the coefficients $c_{n,i}$.
At the level $n=4$ we find no solutions. But adding the $n=6$ terms gives
a one parameter family of solutions and adding $n=8$ an even larger
family of solutions. And this is in spite of the fact we get many more
equations than free parameters.

   A hopeful conclusion is that we have done something right, and the
form of (\ref{cftsol}) is well suited to solution by relatively
lower orders in expected forms of the effective action. But closer
examination of the resulting equations of motion showed that when the
initial condition appropriate to the solution (\ref{cftsol}) are
inserted the resulting equations become degenerate, and we don't
have enough dynamical equations to reconstruct (\ref{cftsol}). In
particular, both of the dynamical equations (\ref{parta}) and
(\ref{partphi}) become $\pbdd(t)=-3 \dot H(t)$.

   Again a hopeful conclusion is that this is evidence for the special
nature of (\ref{cftsol}). But in fact this can be seen to be true
of any $Q=0$ solution. Referring to (\ref{varbeta}), we see that the
conserved quantity for our action ansatz can be written,
\beq
Q=e^{-\bar \phi} \partby{\L}{H},
\eeq
so the $Q=0$ condition becomes $\partby{\L}{H}=0$. So the lower order part
in (\ref{parta}) becomes identically zero. The lower order part in the
second dynamical equation is, in terms of the action (we also set
$n(t)=1$),
\beq
\L-\pbd \partby{\L}{\pbd}=\sum\limits_{n,i} c_{n,i} H(t)^i \pbd(t)^{n-i}
  -\sum\limits_{n,i} (n-i) c_{n,i} H(t)^i \pbd(t)^{n-i}.
\eeq
Combining this with (\ref{partn}),
\beq
\sum\limits_{n,i} (n-1) c_{n,i} H(t)^i \pbd(t)^{n-i}=0,
\eeq
we see,
\beq
\L-\pbd \partby{\L}{\pbd}=\sum\limits_{n,i} i c_{n,i} H(t)^i \pbd(t)^{n-i}
 = H \partby{\L}{H}.
\eeq
So, in fact, the $Q=0$ condition, $\partby{\L}{H}=0$ causes the dynamical
equations to become homogeneous equations in the higher
derivatives. Since a homogeneous system does not have a unique non-zero
solution, we have lost the ability to recover the time dependence of
(\ref{cftsol}) from the equations of motion.

While this disturbing and unexpected, it has another interesting
implication. Any point on the trajectory can be regarded as a fixed 
point, since $(\pbd,H)=constant$ trivially satisfies the eoms 
(\ref{parta}) and (\ref{partphi}) by virtue of the vanishing of the
lower derivative contributions. In other words because of this 
degeneracy of the eoms, instead of having isolated fixed points we have
curves of fixed points.

Furthermore, although the CFT trajectory occupies only a finite segment
in the $(\pbd,H)$ phase space, looking at the situation from the view point
of the eoms this is only part of the story. Consider the equation 
(\ref{partn}) which is just a polynomial in $\pbd$ and $H$. Since it vanishes
on the segment defined by (\ref{cftsol}) it follows that (\ref{partn}) 
contains a linear factor which is just the equation defining the segment,
explicitly $\pbd+3 H(t)+4 K$. Since this linear factor vanishes on the 
entire infinite line containing the segment we find this constrain equation
is valid on the entire line. Similarly, since the quantity
$\partby{\L}{H}=0$ must be satisfied on the segment it must also be 
satisfied on the entire line. So the previous arguments can be seen to 
hold on the entire line. So we have, in fact, have an infinite line of fixed 
points.

Now consider the time reversed solution, clearly it also a $Q=0$ 
trajectory, since when $\pbd \rightarrow -\pbd$ and $H(t) \rightarrow
-H(t)$ we have $\partby{\L}{H} \rightarrow -\partby{\L}{H}$ since it 
only contains terms of odd degree. All of the foregoing apply to it
also, so it can also be extended to an infinite line of fixed points.
We have illustrated these extended lines in \figref{f:fig4}{b}.

\bfig
\bigepstwo{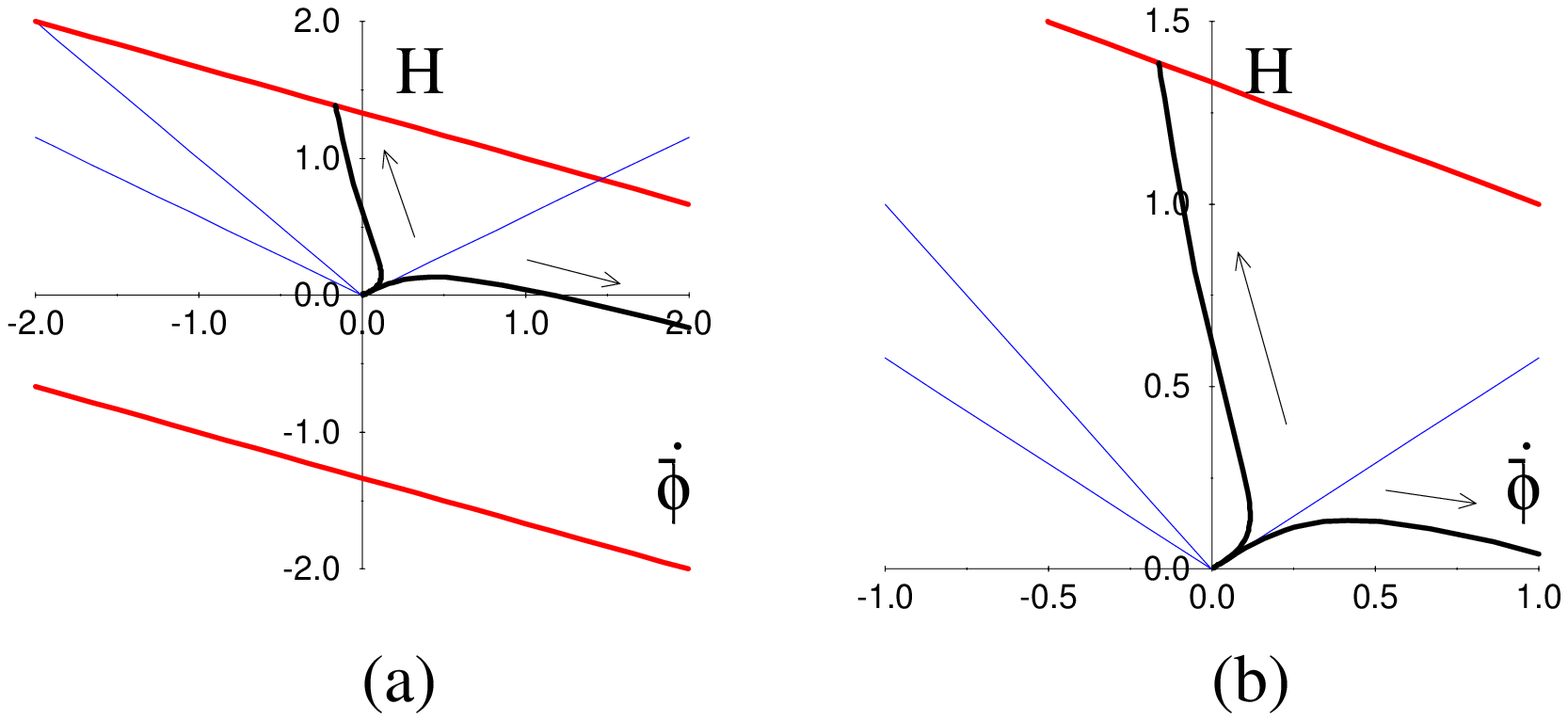}
\caption
{\small\it
a) Trajectories coming from the (+) vacuum, K=1, for effective actions
that contain (\ref{cftsol}) as a solution. The red lines are the lines
of fixed points created by (\ref{cftsol}) (below) and its time reversed
partner (above).
b) A closer view of a) showing the solution going to a fixed point
on the line created by the time reversed partner of (\ref{cftsol}).}
\label{f:fig5}
\efig

Returning our attention to the solution coming from the (+) branch vacuum
where the equations of motion are generically nonsingular, we observe that
it is between two lines of fixed points. The points at $\pbd>0$ are
repulsive and the solution cannot flow into them or cross them. But those
at $\pbd<0$ are attractive and make a large target of future attractors. 
So the (+) branch solution may either flow to infinity or singularity between
the lines or flow to one of the $\pbd<0$ fixed points. The choice between
these alternatives depends on the exact form of the effective action. As
we have stated, requiring (\ref{cftsol}) to be a solution of the effective
action to a given order does not fix the coefficients in the effective 
action but only constrains them. We illustrate the situation in 
\figref{f:fig5}{}. Here we have fitted an effective action of order
$O(\alpha'^2)$ to the CFT solution, which leaves us with one free 
parameter. Setting this parameter to two different values allows us
to exhibit these two different types of behavior by numerically integrating
the (+) branch vacuum solution. While we find
that the solutions are not compelled to flow to the fixed points, such
behavior seems to occur over a large portion of the parameter space.

While these observations depend on the exact nature of the CFT solution 
and we should not expect it to be exact as we are working in conformal
perturbation theory, we remark that much of the previous argument can
be applied to a solution which only qualitatively resembles the CFT solution.
The fact it is a $Q=0$ solution is necessary only because it exits from
a $\pbd<0$ fixed point. As this is the most reliable perturbative regime
(as $t \rightarrow -\infty$), this is perhaps reasonable. And while we 
lose the exact factorization arguments, a line of zeros of a polynomial
expression does not simply terminate at a point as the CFT solution does. 
So we should expect that the resulting curve of fixed points can again
be extended. 

We have made use of the time reversed CFT solution but have made no
mention of SFD. We expect imposing the naive form of SFD will 
only eliminate good fixed point behavior, as we have argued, so we 
do not base any arguments on it. We do expect that the form
of the action should reflect some form of SFD, but without knowing
something of its nature it is difficult to be exact. Since the naive
SFD partner of the CFT solution flows out of a $\pbd<0$ fixed point we
should expect the exact SFD partner does as well, making it in turn
a $Q=0$ solution and a line of fixed points. So we may find other walls
of fixed points around as well, compelling the solution coming out of 
the (+) vacuum to have good fixed point behavior.

\renewcommand{\theequation}{3.\arabic{equation}}
\setcounter{equation}{0}
\section {Conclusions}

We have seen that the known information about the nature of the
$O(\alpha')$ corrections to the effective action coming from string
theory are insufficient to decide whether inflationary branch used
in the pre-big-bang scenario exhibits curvature saturation by 
flowing into a fixed point. Attempts 
to use SFD to further constrain the action finally lead an exact statement
independent of the order of correction that naive SFD 
simply works against this behavior. However we
have also observed that naive SFD simply cannot be implemented at
$O(\alpha')$ in the general anisotropic case and we concur with other
work suggesting that SFD itself must receive higher order corrections. 

We have also scanned the parameter space of possible forms of corrections
and have determined that good fixed point behavior, while not universal,
does occupy a large region of this parameter space. Finally, we have 
attempted to construct a plausible approximation to a conformally exact
solution and discovered that independent of its exact form, a generically
similar solution forces the equations of motion to 
a corresponding action to exhibit a degeneracy which forces the 
existence of continuous lines of fixed points. These fixed points can
in turn powerfully constraint the possible evolution of the (+) 
inflationary branch, opening the possibility that deeper knowledge of
some conformally exact solutions may be enough to settle the question
of whether string theory predicts curvature saturation for the inflationary
scenario of string cosmology.

\vskip 2 cm
\noindent
{\Large\bf Acknowledgment}\\

This research is supported in part by the Israel Science Foundation 
administered by the Israel Academy of Sciences and Humanities.

\end{document}